\documentclass{vgtc}                          

\graphicspath{{figures/}{pictures/}{images/}{./}} 

\usepackage{times}                     

\usepackage{tabu}                      
\usepackage{booktabs}                  
\usepackage{lipsum}                    
\usepackage{mwe}                       

\usepackage{mathptmx}                  
\usepackage{amsmath}
\usepackage{amsfonts}
\usepackage{subcaption}

\onlineid{0}

\vgtccategory{Research}

\vgtcinsertpkg

\title{Glyph-Based Uncertainty Visualization and Analysis of Time-Varying Vector Fields}

\author{%
 Timbwaoga A. J. Ouermi \thanks{e-mail: touermi@sci.utah.edu} \\
 \scriptsize SCI Institute, University of Utah
 \and Jixian Li \thanks{e-mail: jixianli@sci.utah.edu} \\
 \scriptsize SCI Institute, University of Utah
 \and Zachary Morrow \thanks{e-mail: zbmorro@sandia.gov} \\
 \scriptsize Sandia National Laboratories
 \and Bart van Bloemen Waanders \thanks{e-mail: bartv@sandia.gov} \\
 \scriptsize Sandia National Laboratories
 \and Chris R. Johnson \thanks{e-mail: crj@sci.utah.edu} \\ 
 \scriptsize SCI Institute, University of Utah
}

\teaser{
  \centering
  \includegraphics[width=0.75 \linewidth]{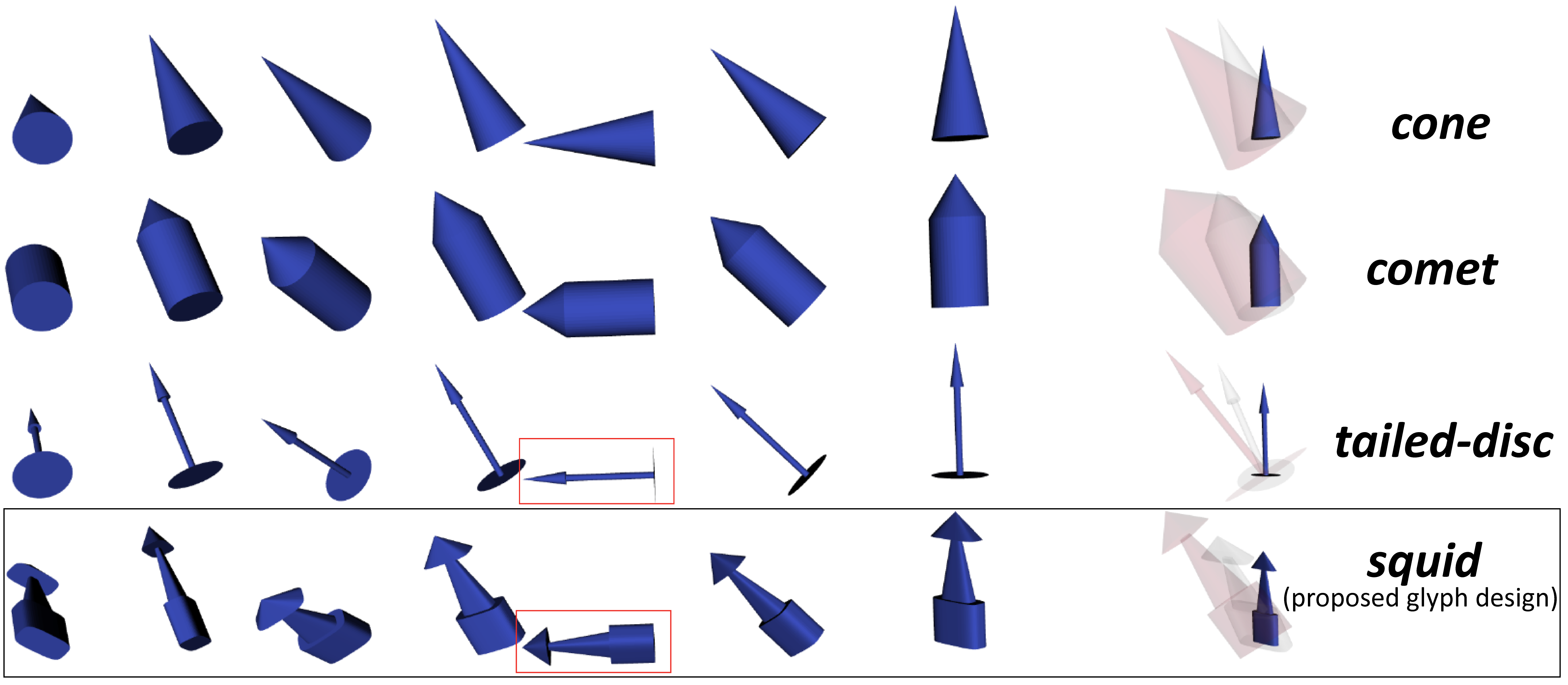}
  \vspace{-2.5mm}
  \caption{3D vector uncertainty glyph. The glyphs' direction corresponds to the median vector direction. The angle and length of the \textit{cone} glyph represent the angular variation (the maximum angle between all pairs of vectors within the ensemble) and the maximum vector length. The magnitude variation (the difference between the maximum and minimum magnitude) is not included. The \textit{comet} glyph encodes the magnitude variation and minimum magnitude using the length of the cone and cylinder, respectively. However, these variations are not easily discernible. While both the \textit{tailed-disc} and \textit{squid} clearly distinguish these uncertainties, the small arrow size and rotational symmetry of the \textit{tailed-disc} limit the perception of the size of the uncertainty. In contrast, our proposed \textit{squid} glyph effectively distinguishes between magnitude and direction variations. Additionally, it employs a superellipse (2D superquadric) to better approximate directional variations, eliminating rotational ambiguity around the median direction. The last column shows the uncertainty variation between the selected time step (opaque blue) and the next two time steps shown in transparent grey and red. The \textit{squid} glyph is more accurate and less prone to rotational degeneracy.}
  \label{fig:teaser}
}

\abstract{
   Uncertainty is inherent to most data, including vector field data, yet it is often omitted in visualizations and representations. Effective uncertainty visualization can enhance the understanding and interpretability of vector field data. For instance, in the context of severe weather events such as hurricanes and wildfires, effective uncertainty visualization can provide crucial insights about fire spread or hurricane behavior and aid in resource management and risk mitigation. Glyphs are commonly used for representing vector uncertainty but are often limited to 2D. In this work, we present a glyph-based technique for accurately representing 3D vector uncertainty and a comprehensive framework for visualization, exploration, and analysis using our new glyphs. We employ hurricane and wildfire examples to demonstrate the efficacy of our glyph design and visualization tool in conveying vector field uncertainty. 
} 


\begin{document}


\firstsection{Introduction}

\maketitle
Visualization and analysis of vector field data is fundamental to many scientific applications including flow simulations, weather prediction, and wildfire simulations. For example, in the context of wildfires, understanding wind uncertainty and flow patterns is critical for forecasting the overall fire behavior and ultimately providing more information to aid in resource management and risk mitigation. The wind data in these scenarios are inherently time-varying, three-dimensional, and associated with uncertainties. 

Effective techniques for visualizing uncertainty in vector field data can improve understanding and interpretability ~\cite{Lie2009, Ward2008, Borgo2013}. Uncertainty visualization provides an understanding of the limitations and potential errors in the data, as well as insight into the observed features and trends in the vector field. 

In this work, we present a glyph-based method for visualization and analysis of time-dependent 3D vector field uncertainty. Glyph-based methods for vector field uncertainty visualization have mostly been limited to 2D ~\cite{Post1995, Boring1996, Wittenbrink1996, Zuk2008, Hlawatsch2011, Jarema2015, Wiebel2012, Borgo2013, Chung2014}, and existing 3D glyph-based methods are prone to rotational ambiguity and do not provide insight into the distribution of vectors within the high uncertainty region. 
We address these limitations through the following contributions:
   (1) A 3D uncertainty glyph, we have named the \textit{squid} glyph, that is designed using superellipses and cones to provide a more accurate approximation of the vector magnitude and direction uncertainty compared to the uncertainty \textit{cone}~\cite{Lee2018}, \textit{comet} (cylinder + cone)~\cite{McQuinn2014}, and \textit{tailed-disc} (disc + arrow)~\cite{Lee2018} glyphs. Our \textit{squid} glyph is suitable for representing anisotropy in directional variation and is less prone to visual ambiguity;
   (2) The use of multivariate data depth, which we refer to as ``vector depth'' for exploration and analysis. The vector depth measures how immersed a given vector is compared to other vectors in a set. This metric provides insight into the distribution and individual vectors at each spatiotemporal location without prior assumptions about the distribution; and
   (3) A visualization framework
   that utilizes the \textit{squid} glyph, the vector depth, and other methods to enable interactive visualization, exploration, and analysis of time-varying vector field uncertainty. 

%
\section{Related Work}
\label{sec:related-work}
Several glyph-based methods have been developed to integrate uncertainty in order to improve understanding and interpretability of vector field visualization~\cite{Wittenbrink1996, pang1997, schmidt2004, Hlawatsch2011, Jarema2015, Lee2018, Zuk2006, Johnson2003, Zuk2008, Borgo2013}. Uncertainty in vector fields is often incorporated into the visualization as an add-on by overloading or modifying visual attributes such as surface reflectance, rendering opacity, and color to represent the uncertainty~\cite{Bonneau2014, Boring1996, pang1997}. 

The work in~\cite{Wittenbrink1996, Wittenbrink1995, pang1997} proposes different 2D glyph designs where the vector magnitude and angular variation are encoded in the length and shape of the glyph. Jarema et al.~\cite{Jarema2015} propose a glyph-based methods 
that fits the vector field distribution at each location with the Gaussian mixture model (GMM) and a 2D lobular glyph overloaded with a colormap to visualize the distribution and uncertainty. Halwatsch et al.~\cite{Hlawatsch2011} proposed flow-radar glyphs for visualization of time-varying vector fields and their extension to incorporate magnitude and direction uncertainty in 2D. The flow-radar glyph is constructed by spatially and radially organizing the temporal vector direction into a sequence of points that are connected to form the glyph. Zuk et al.~\cite{Zuk2008} utilize animation with 2D glyph representations to visualize the bidirectional vector field uncertainty in the context of analyzing anisotropy of rocks.  Lee and Park~\cite{Lee2018} proposed a 3D glyph design to improve the representation of uncertainty and visual perception of the vector field data. This approach combines a disc and arrow glyph to encode the uncertainty. Schmidt et al.~\cite{schmidt2004} use box, sphere, and cylinder glyphs to represent scalar uncertainty and arrow glyphs overloaded with color to represent directional uncertainty. 
Post et. al~\cite{Post1995} proposed geometric primitives such as arrows and ellipses to encode and render 3D vectors and its distribution.


%
%
\section{Methods}
\label{sec:methods}
Several uncertainties, including measurement, data transformation, and parameter variation, occur in 3D vector field. The instruments typically provide the uncertainty for measured data ~\cite{Wilczak1995}. The uncertainties from data transformation and parameter variations can be estimated from ensemble data with statistical methods. \Cref{sec:related-work} highlights different glyph-based methods for encoding the estimated uncertainties to different glyph attributes. 
\subsection{Uncertainty Glyph Design}
\label{subsec:uncertainty-glyph}
\begin{figure}[!ht]
    \centering
    \includegraphics[width=0.8\columnwidth]{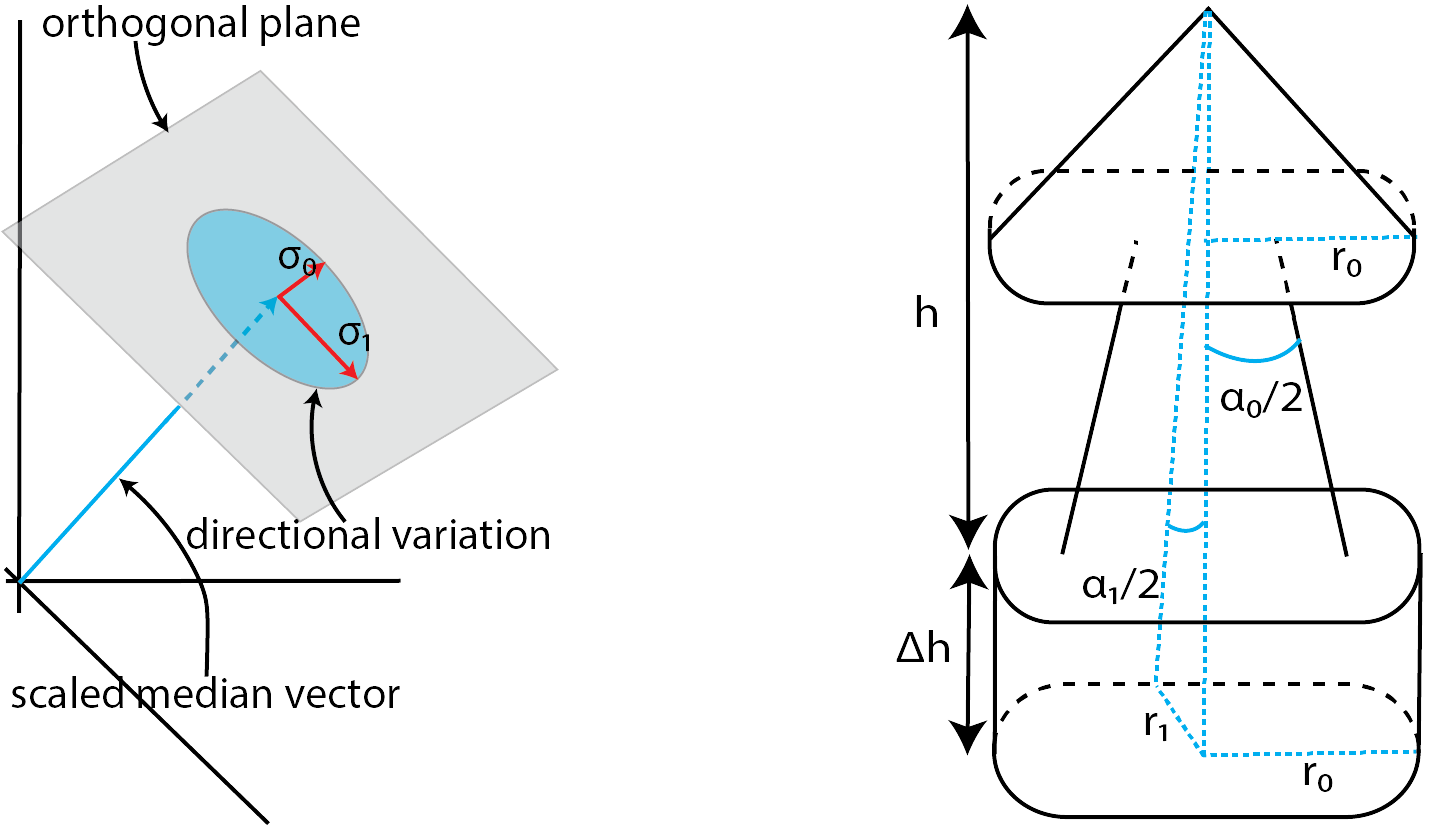}
    \vspace{-2.5mm}
    \caption{Uncertainty \textit{squid} glyph design: The visualization on the left side illustrates calculating the first ($\sigma_{0}$) and second  ($\sigma_{1}$) PCA principal components used to represent the directional spread and scale of the superellipse semi-minor axis. The diagram on the right side shows the \textit{squid} glyph design with its parameters. 
    The magnitude and magnitude variation are represented along the glyph direction. The minimum magnitude, magnitude variation, and maximum magnitude are represented with $h$, $\Delta h$, $h + \Delta h$, respectively. 
    }
    \label{fig:uncertainty-glyph}
\end{figure}

The uncertainty glyph design is crucial for effectively visualizing the vector field uncertainty. We introduce the \textit{squid} glyph, a 3D uncertainty glyph that builds on the ideas introduced in ~\cite{Wittenbrink1996, pang1997} and leverages the design guidelines in ~\cite{Ward2008, Lie2009, Borgo2013}. Our \textit{squid} glyph design primarily focuses on (1) accurately encoding the magnitude and direction uncertainty, (2) ensuring that the information encoded in the glyph is easily discernible, and (3) utilizing intuitive representations to provide insight into the vector field patterns and their uncertainties. We discuss these considerations sequentially. The design details of our \textit{squid} glyph are illustrated in \cref{fig:uncertainty-glyph}.

First, accurately mapping uncertainty information to glyph attributes is essential for the correct interpretation of individual uncertainty glyph and their relationships to neighboring glyphs. 
The use of ellipses can improve accuracy; however, it introduces a rotational degeneracy about the median direction of the glyph.  In tensor visualization, superquadrics have been employed to enhance accuracy and remove rotational ambiguity ~\cite{Gordon2004,Thomas}. Following a similar approach, we utilize superellipses to design an uncertainty glyph that is both more accurate and less susceptible to rotational ambiguity. In \cref{fig:uncertainty-glyph}, 
the minimum magnitude $h$, magnitude variation $\Delta h$, and the maximum magnitude $h + \Delta h$  are mapped to the length of the body, head, and entire glyph, respectively. The directional uncertainty is mapped to the angles $\alpha_{0}$, $\alpha_{1}$ and the semi-major axis $r_{0}$ and semi-minor axis $r_{1}$ of the superellipse at the base of the \textit{squid} glyph shown on the right side of \cref{fig:uncertainty-glyph}. The semi-major and semi-minor axis indicate the spread of the directional variation is calculated according to 
\begin{equation*}
    r_{0} = (h + \Delta h) \tan \big( \frac{1}{2}\alpha_{0} \big), \quad r_{1} = r_{0}\frac{||\sigma_{1}||}{||\sigma_{0}||},  \quad \alpha_{1} = \tan^{-1} \left( \frac{h + \Delta h}{r_{1}} \right)
\end{equation*}
where $\alpha_{0}$ is the maximum angle from the ensemble vectors at each spatiotemporal location. The visualization on the left side of \cref{fig:uncertainty-glyph} illustrates the process for calculating $\sigma_{0}$ and $\sigma_{1}$. They are obtained by finding the intersection of each vector with a plane orthogonal to the scaled median vector $(||v_\text{max}||/||v_\text{median}||)v_\text{median}$ and crosses its tip. 
The median vector $v_\text{median}$ corresponds to the vector with the largest ``vector depth" which is multivariate data depth that will be described in \cref{subsec:data-depth}. 
We then employ principal component analysis (PCA) decomposition to find the first and second principal directions $\sigma_{0}$ and $\sigma_{1}$, respectively. The ``squid" glyph direction corresponds to the median direction.

Second, individual glyph parts encoding the data information should be easily discernible. To facilitate interpretability, the glyph parts should be ``orthogonal,'' meaning that each glyph part should be visually distinguishable independently~\cite{Lie2009}. To ensure orthogonality, our uncertainty \textit{squid} glyph is designed such that the head, body, magnitudes, and angles can be perceived separately. These glyph attributes are normalized across the vector field data to enable comparison based on glyph sizes.

Lastly, given that arrows are a common and intuitive representation of directions and vectors, we utilize the cone and super-ellipse primitives to construct an arrow-like glyph that intuitively indicates size and direction. Glyphs that are designed based on semantic meaning are well-suited for interpretability and identifying features and patterns for visualization~\cite{Ward2008, Lie2009, Borgo2013}. 

Occlusion and depth-perception limitations in 3D visualization impede the user's ability to effectively interpret spatial relationships and structures of the vector field and its uncertainty. We address this limitation by providing global and local interactive visualization capabilities that enable the user to focus on specific regions, or slices, thereby reducing occlusion and limitations of depth perceptions. 

\Cref{fig:teaser} shows a comparison of our uncertainty \textit{squid} glyph with the \textit{comet}, \textit{tailed-disc}~\cite{Lee2018}, and the standard \textit{cone} glyphs. The \textit{cone} glyph only represents angular variation and not magnitude variations. The \textit{comet} glyph does not clearly distinguish magnitude and direction variations. While both \textit{tailed-disc} and \textit{squid} glyphs depict these uncertainties, the \textit{tailed-disc}'s small arrow size and rotational symmetry around the median direction limit the perception of the glyph's size, as highlighted in the orange boxes. In contrast, the \textit{squid} glyph design effectively distinguishes between magnitude and direction variations. Additionally, it employs a superellipse to better approximate directional variations, eliminating rotational ambiguity around the median direction. 
Overall, the \textit{squid} glyph is more accurate and less prone to rotational degeneracy.

\subsection{Vector Depth}
\label{subsec:data-depth}
``Vector depth'' is a multivariate data depth, a non-parametric statistical method for measuring centrality. It measures how deep or immersed a given vector is compared to other vectors in the dataset. The \textit{squid} glyph in \cref{subsec:uncertainty-glyph} provides a summary of the uncertainty at each spatiotemporal location, while the vector depth offers insights into the distribution of vectors from which the uncertainty summary is derived. The multivariate data depth approaches ~\cite{Regina1999, Barnett1976, regina1988} have been extended to characterize contours ~\cite{whitaker2013}, and curves ~\cite{Mirzaragar2014}. Here, we adapt the multivariate data depth to vectors by computing the vector depth in spherical coordinates instead of standard Cartesian coordinates. Spherical coordinates are better suited for representing magnitude and direction variations. 

Let $F$ be a continuous probability distribution in $\Omega=\mathbb{R}\times[0, \pi]\times[-\pi,\pi]$, with $V = \{ X_{1}, \dots, X_{n}\} \subset \Omega$ a collection of $n > 3$ random samples from $F$. 
Any point in the 3D Cartesian coordinate ($\mathbb{R}^{3}$) can be uniquely represented in the spherical coordinate ($\Omega$). 
The vector depth of $x$ is the percentage of size-$4$ subsets of $V$ whose bounding box in $\Omega$ contains $x$. Mathematically,
\begin{equation}
    VD(x) = \binom{n}{4}^{-1} \sum_{\tilde{V} \subset V,~|\tilde{V}|=4} \mathbf{1}_{S[\tilde{V}]}(x). 
\end{equation}
$S[\tilde{V}]$ is the closed region in spherical space with corners defined by $\tilde{V}$, and $\mathbf{1}$ is the indicator function (1 if $x \in S[\tilde{V}]$, 0 otherwise). 

\begin{figure}[!ht]
    \centering
    \includegraphics[width=0.4\columnwidth]{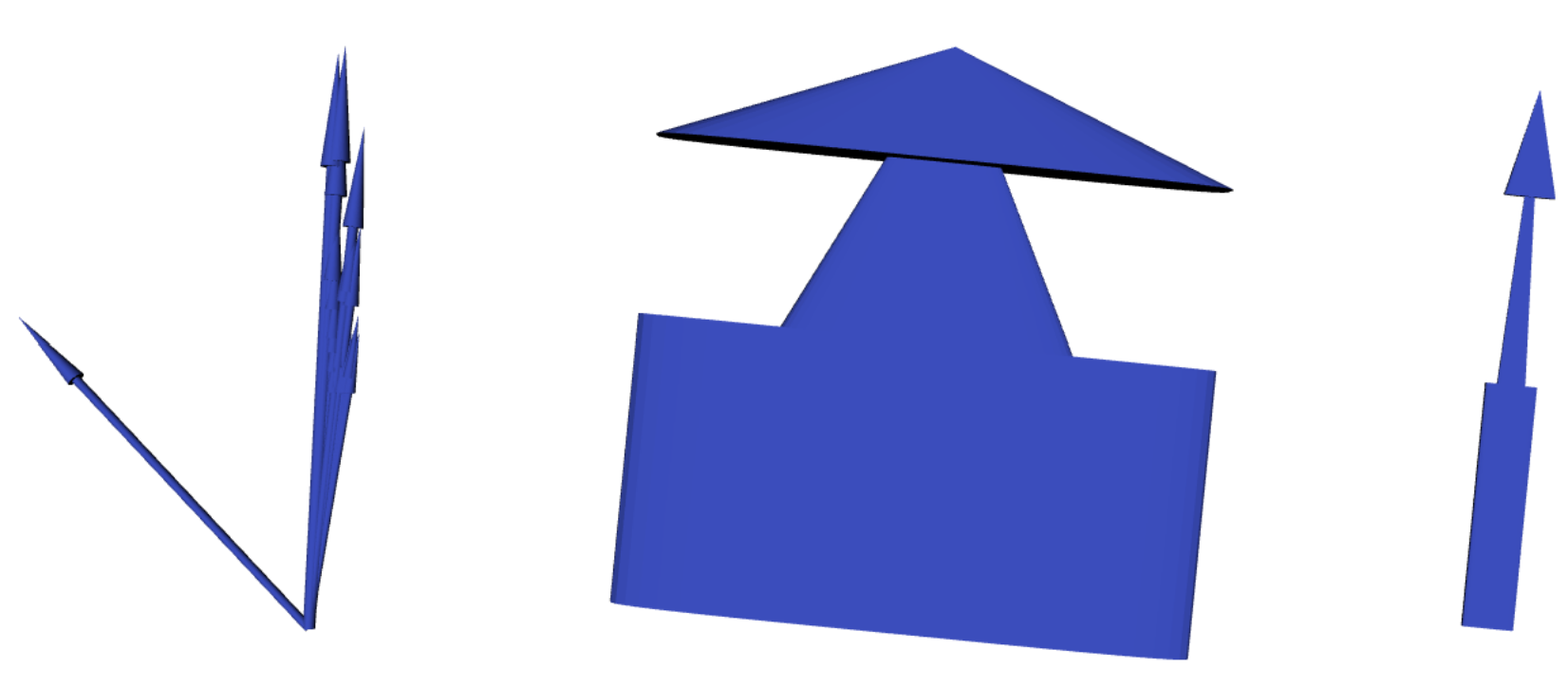}
    \vspace{-2.5mm}
    \caption{The arrows on the left side show the explicit visualization of the vector distribution. The \textit{squid} glyph approximation for this distribution is shown in the center. The \textit{squid} glyph representation without the outlier is shown on the left side. }
    \label{fig:vector-depth}
\end{figure}

The data depth is suitable for identifying outliers and providing insight into vector distribution. \Cref{fig:vector-depth} shows a skewed distribution with an outlier on the left side, a \textit{squid} glyph approximation of the distribution in the center, and a \textit{squid} approximation without the outlier that is removed by excluding the smallest vector. 
\subsection{Vector Field Uncertainty Visualization Tool}
\label{subsec:vector-uncertainty-analysis-framework}
We present a visualization tool for the exploration and analysis of time-varying vector field data. Vector field uncertainties are often omitted, and when included, they are typically limited to 2D representations. Our framework leverages color, transparency, multiple views, the \textit{squid} glyph, and ``vector depth'' of the previous two sections to enable a comprehensive global and local spatiotemporal analysis of vector-field uncertainty. The interface is designed to highlight and provide insights into regimes with significant uncertainty while preserving the overall visual patterns of the vector field.

We employ an example dataset constructed by sampling the vector field in \cref{eq:synthetic-example} and adding uniform random noise to each vector component. Specifically, the generated dataset includes $20$ ensemble members at each spatiotemporal location, obtained by perturbing each vector component with uniform noise.
\begin{equation}\label{eq:synthetic-example}
    \begin{matrix}
        u_{0} = \sin(x) \\ 
        v_{0} = \sin(y) \\
        w_{0} = 0.5
    \end{matrix}
    \quad
    \begin{bmatrix}
        u_{t+1} \\
        v_{t+1} \\
        w_{t+1} 
    \end{bmatrix}
    =
     \begin{bmatrix}
        u_{t} \cos(t) - v_{t} \sin(t) \\
        u_{t} \sin(t) + v_{t} \cos(t) \\
        w_{t}
    \end{bmatrix} 
    \quad
    \begin{matrix}
      x \in [-1, 1] \\
      y \in [-1, 1] \\
      t \in [0, \frac{3}{4}\pi].
    \end{matrix}
\end{equation}

\begin{figure}[!ht]
    \centering
    \includegraphics[width=0.99\columnwidth]{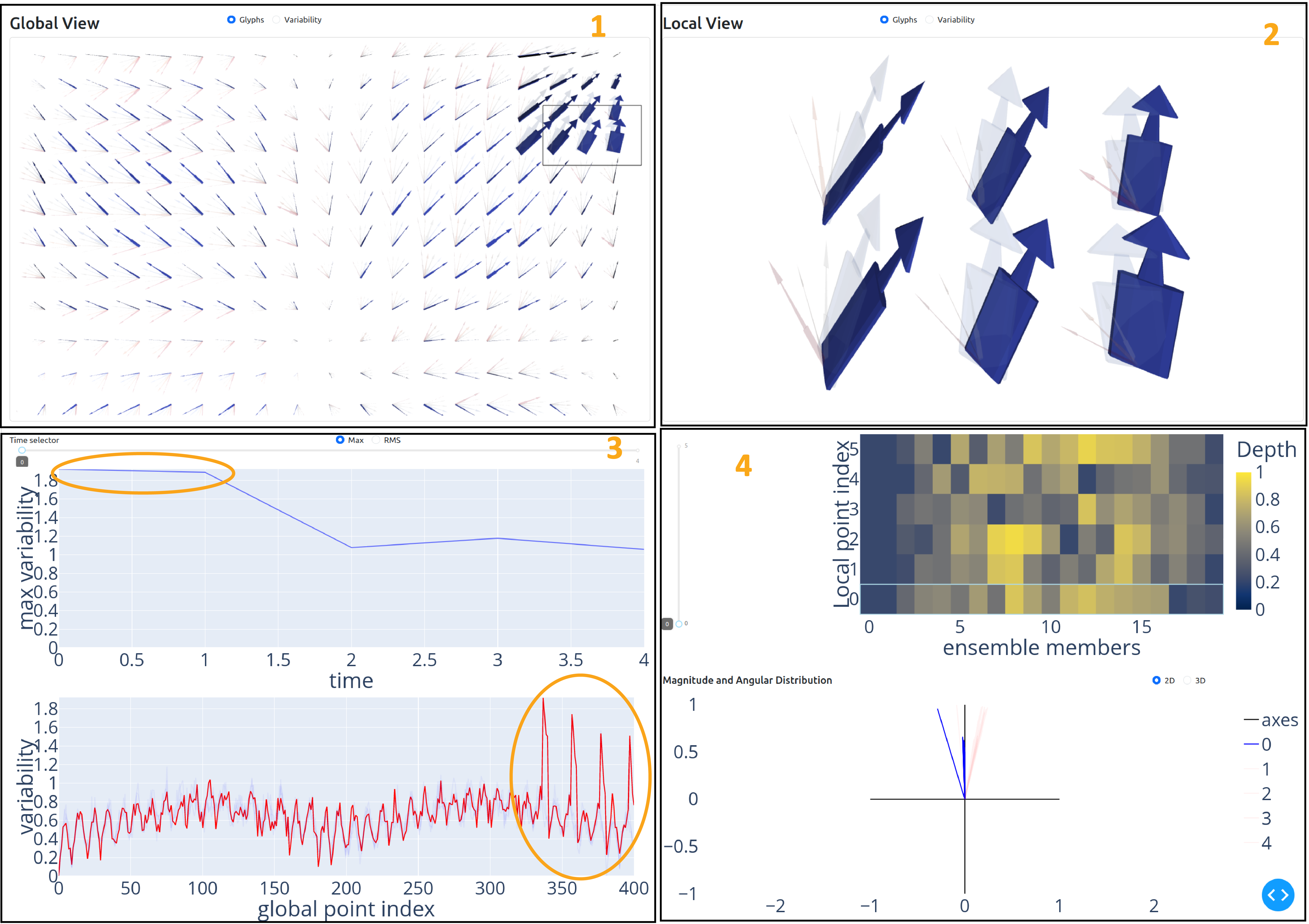}
    \vspace{-2.5mm}
    \caption{Vector uncertainty analysis interface. The outlined black rectangles and orange numbers indicate the components. 
    $C_{1}$ and $C_{2}$ are used for global and local visualization of filtered vector data, respectively. $C_{3}$ visualizes the magnitude variations and while $C_{4}$ visualizes the vector depth distribution magnitudes and angular variations at a selected location. 
    }
    \label{fig:interface}
\end{figure}

\Cref{fig:interface} shows the framework interface with the components in black rectangles and the numbers in orange. 
%

\begin{figure*}[!ht]
    \centering
    \vspace{-3mm}
    \begin{subfigure}[b]{0.16 \textwidth}
        \includegraphics[width=\columnwidth]{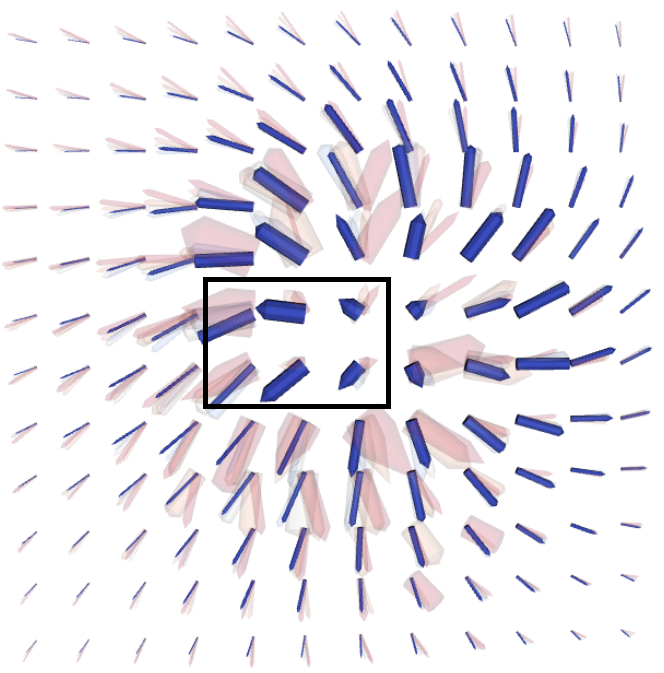}
        \label{subfig:hurricane_comet_local}
    \end{subfigure}
     \vspace{-2mm}
    \begin{subfigure}[b]{0.16 \textwidth}
        \includegraphics[width=\columnwidth]{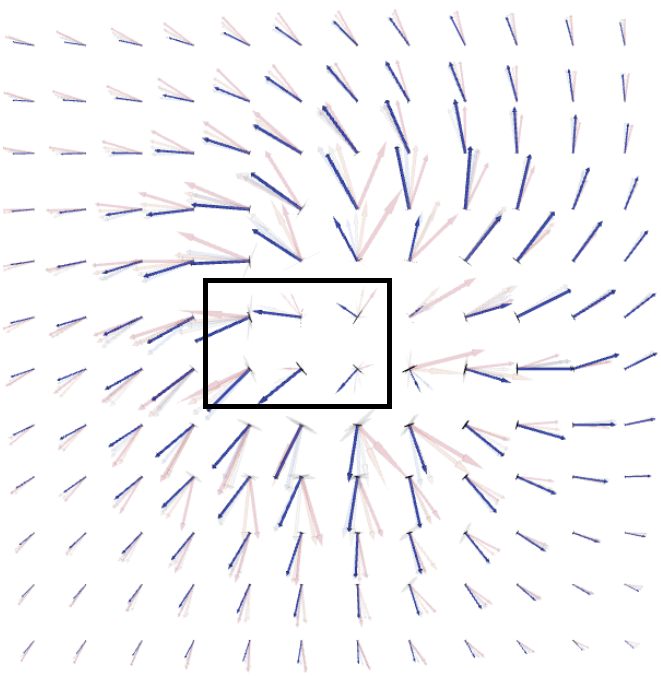}
        \label{subfig:hurricane_disc_arrow_local}
    \end{subfigure}
    \begin{subfigure}[b]{0.16 \textwidth}
        \includegraphics[width=\columnwidth]{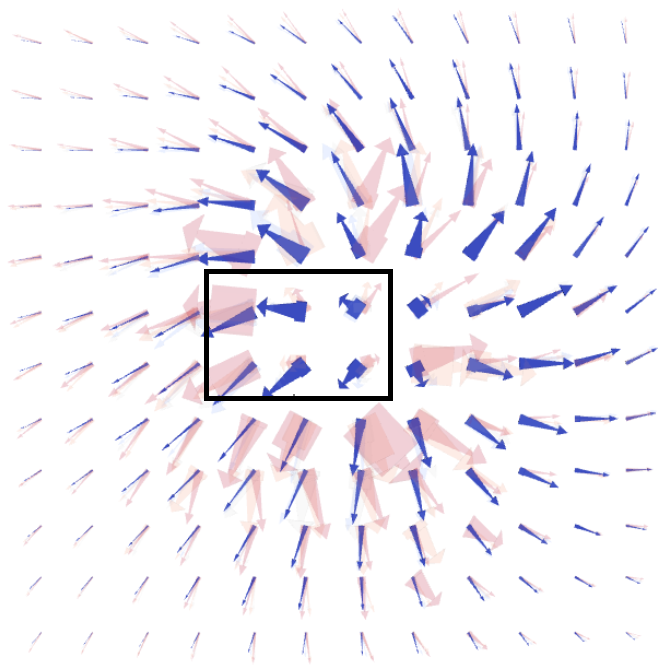}
        \label{subfig:hurricane_squid_local}
    \end{subfigure}
    \begin{subfigure}[b]{0.16 \textwidth}
        \includegraphics[width=\columnwidth]{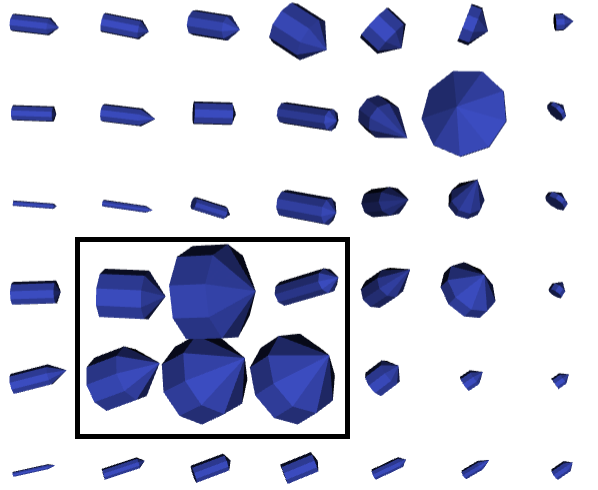}
        \label{subfig:fire_wind_comets_local}
    \end{subfigure}
    \begin{subfigure}[b]{0.16 \textwidth}
        \includegraphics[width=\columnwidth]{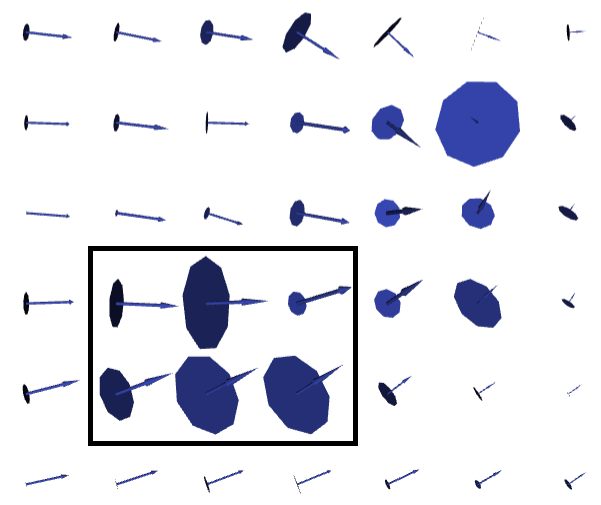}
        \label{subfig:fire_wind_disc_arrows_local}
    \end{subfigure}
    \begin{subfigure}[b]{0.16 \textwidth}
        \includegraphics[width=\columnwidth]{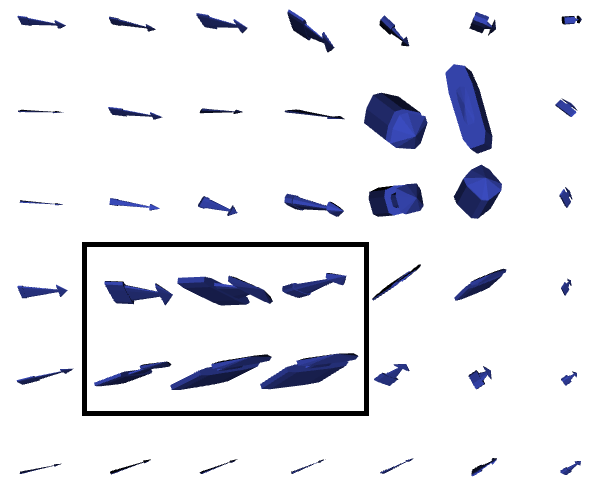}
        \label{subfig:fire_wind_squids_local}
    \end{subfigure}
    \begin{subfigure}[b]{0.16 \textwidth}
        \includegraphics[width=\columnwidth]{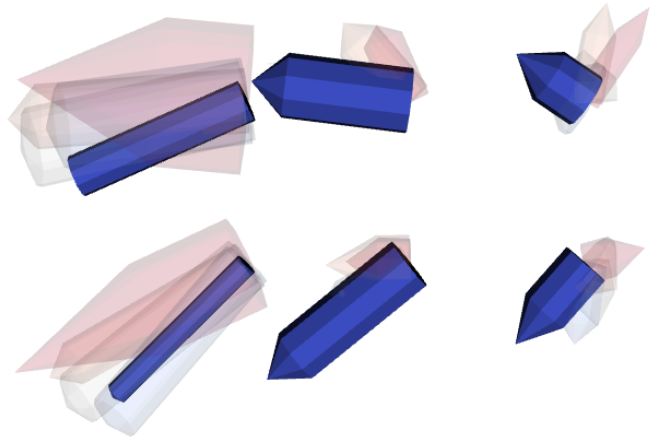}
        \caption{\textit{comet} (hurricane)}
        \label{subfig:hurricane_comet_local_zoom}
    \end{subfigure}
    \begin{subfigure}[b]{0.16 \textwidth}
        \includegraphics[width=\columnwidth]{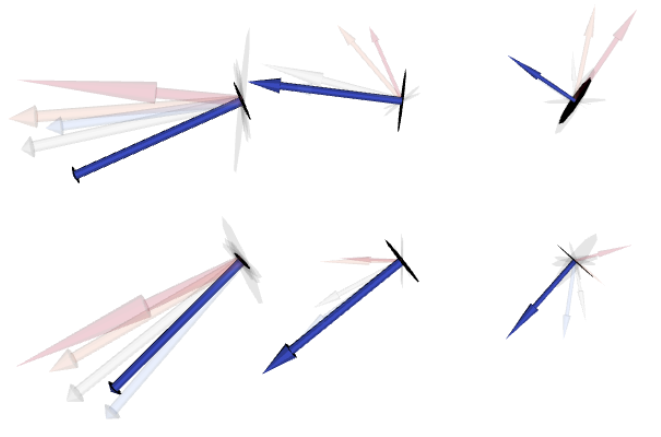}
        \caption{\textit{tailed-disc}  (hurricane)}
        \label{subfig:hurricane_disc_arrow_local_zoom}
    \end{subfigure}
    \begin{subfigure}[b]{0.16 \textwidth}
        \includegraphics[width=\columnwidth]{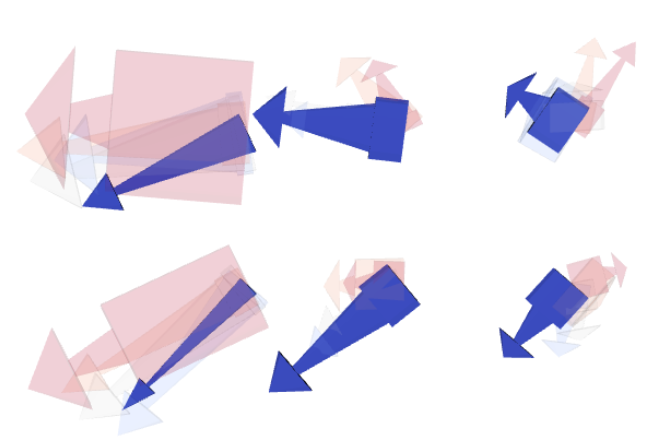}
        \caption{\textit{squid} (hurricane)}
        \label{subfig:hurricane_squid_local_zoom}
    \end{subfigure}
    \begin{subfigure}[b]{0.16 \textwidth}
        \includegraphics[width=\columnwidth]{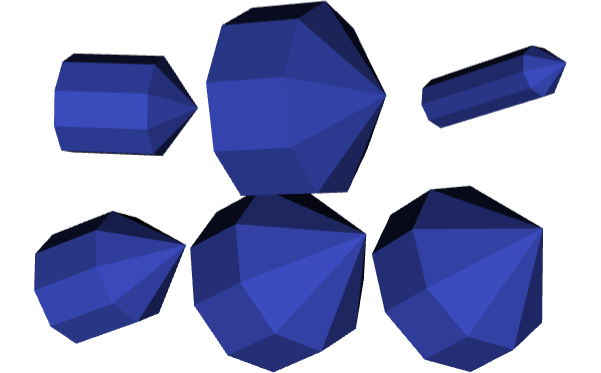}
        \caption{\textit{comet} (wildfire)}
        \label{subfig:fire_wind_comets_local_zoom}
    \end{subfigure}
    \begin{subfigure}[b]{0.16 \textwidth}
        \includegraphics[width=\columnwidth]{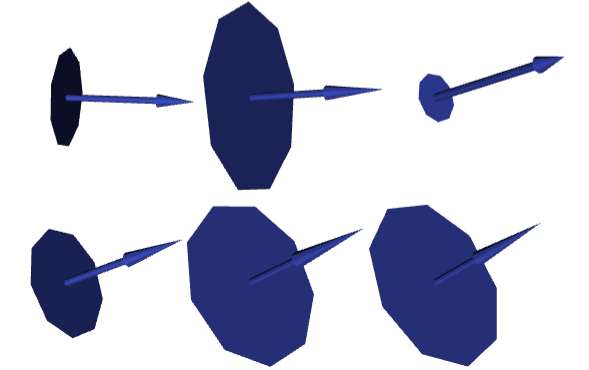}
        \caption{\textit{tailed-disc} (wildfire)}
        \label{subfig:fire_wind_disc_arrows_local_zoom}
    \end{subfigure}
    \begin{subfigure}[b]{0.16 \textwidth}
        \includegraphics[width=\columnwidth]{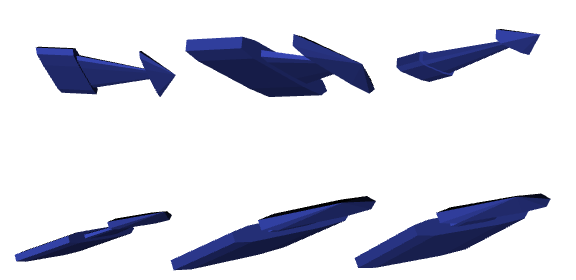}
        \caption{\textit{squid} (wildfire)}
        \label{subfig:fire_wind_squids_local_zoom}
    \end{subfigure}
    \vspace{-2.5mm}
    \caption{Hurricane Isabel and wildfire examples. Both examples show a comparison of the \textit{comet}, \textit{tailed-disc} and \textit{squid} glyphs}
    \label{fig:fire_wind_example}
\end{figure*}

The component $C_{1}$(top left) uses the \textit{squid} glyph to show a global visualization of the vector field data while the second component $C_{2}$ (top right) shows a local visualization based on the selected region indicated by the black box in $C_{1}$. 
The component $C_{3}$ (bottom left) enables the visualization of magnitude variations for the specific a time range and specific selected time. The top plots in $C_{3}$ of \cref{fig:interface} show the maximum magnitude variations (on the y-axis) versus time (x-axis) and the bottom plot shows the magnitude variations at each location for a selected time step. Both visualizations can help the user identify regions with large variations as shown with orange ellipses.The component $C_{4}$ visualizes the vector depth values and provides insight about the distribution of vectors at a selected spatiotemporal point. The top figure in $C_{4}$ uses heatmap visualization where the y-axis represents the spacial locations, the $x$-axis represents the ensemble members, and the colormap represents vector depth for the vectors in the $C_{2}$ at a selected time step. The bottom figure provides detailed information about the distribution of vectors from which the vector uncertainty is computed. The angular variation is calculated with respect to the median vector direction.  These visualizations provide detailed information about the vector field uncertainty and can be employed to detect outliers, as shown in the bottom plot in $C_{4}$.

The variations are computed with respect to the data to provide an intuitive visualization and uncertainty analysis at the measured or sampled spatiotemporal locations directly. These computed variations can be used to guide measurement and resampling offline to reduce uncertainty and improve data quality. In future studies, we plan to investigate Combining spatial, temporal, and parameter-based uncertainty altogether to provide additional insight and further enhance the uncertainty analysis.

\section{Results and Discussion}
\label{sec:results}
\subsection{Hurricane Isabel}
\label{subsec:hurricane}
Hurricane Isabel was the most costly hurricane of the 2003 season with \$3.6 billion in damage. In the context of hurricanes, understanding wind patterns and uncertainty is crucial for a better characterization of the hurricane behavior and to provide useful information that can aid in resource management and risk mitigation. The visualization dataset is obtained by down-sampling the original dataset ~\cite{jiang2004visualization} volume resolution from $500 \times 500 \times 100 \times 48$ to $40 \times 40 \times 100 \times 48$. The vector uncertainty at each sub-sampled location is obtained using the local neighborhood vectors in the $5 \times 5 \times 1$ spatial patch. 

\Cref{subfig:hurricane_comet_local_zoom} - \Cref{subfig:hurricane_squid_local_zoom} show a comparison of the different glyph representations around the hurricane eye for a horizontal slice. The larger-size \textit{comet} glyphs in  \cref{subfig:hurricane_comet_local_zoom} better highlight the regions with large uncertainty compared to the other approaches. However, the \textit{comet} glyphs in \cref{subfig:hurricane_comet_local_zoom} do not effectively differentiate between magnitude and directional variations. In contrast, these variations are distinguished in \textit{tailed-disc} glyphs shown in \cref{subfig:hurricane_disc_arrow_local_zoom}, but the \textit{tailed-disc} design is not suitable for highlighting regions with large uncertainties. \Cref{subfig:hurricane_squid_local_zoom} show that the proposed \textit{squid} glyph design is suitable for (1) identifying regions with larger uncertainties, (2) distinguishing the magnitude and directional variation in each glyph, and (3) more accurately encoding the directional uncertainty dispersion compared the other approaches. The zoomed-in region enables uncertainty comparison of the currently selected time step with neighboring time steps indicated by using transparency and different colors. 
These results demonstrate that the \textit{squid} glyph provides a more accurate representation of the vector uncertainties compared to the \textit{comet} and \textit{tailed-disc}. 

\subsection{Wildfire}
This example models wildfire-produced winds resulting from uncertain fuel properties in the Valles Caldera National Preserve of northern New Mexico. This location was selected for its varied topography, well-defined fuel regions, and history of fire activity. There are three easily identifiable fuel regions: the caldera floor, forest edge, and forest. In reality, fuels are strongly dependent on fine-grained local properties, but state-of-the-art models treat them as bulk quantities for computational tractability, resulting in epistemic uncertainty. Moreover, material parameters in fire codes are typically set to {\em a priori} nominal values based on standard fuel models, e.g.~\cite{anderson1982aids, scott2005standard}, resulting in aleatoric uncertainty.

For each of the three fuel regions, we sample the surface-area-to-volume ratio (SAVR) from a uniform distribution centered on the nominal parameter value. Using the state-of-the-art code WRF-SFIRE~\cite{wrf_sfire}, we run a fire simulation for each SAVR sample, lasting $t=10$ hours. The quantity of interest is the final-time, three-dimensional wind, defined on a $69 \times 53$ grid where $\Delta x = \Delta y = 150$m. The winds are initialized modestly at 1 m/s for dynamical stability, and topographic data comes from LANDFIRE~\cite{landfire}.

\Cref{subfig:fire_wind_comets_local_zoom} - \Cref{subfig:fire_wind_squids_local_zoom} show a comparison of the different glyph designs applied to wildfire winds. The results in the zoomed-in region show that all four glyphs effectively highlight the region with large uncertainty shown in the black box. The magnitude and directional variations are more distinguishable in both the \textit{tailed-disc} and \textit{squid} glyphs, compared to the \textit{comet} glyphs where the different parts are not easily discernible. Moreover, the \textit{squid} glyph provides a more accurate representation and superior visualization of wind uncertainty. Additionally, the \textit{squid} glyph provides a more accurate representation of the directional variations compared to the other methods that assume symmetry in the directional spread around the median direction. This provides insights into overall wind patterns and their associated variations due to changes in fuel content, thereby enhancing understanding of wildfire behavior.

\section{Conclusion}
\label{sec:conclusion}
In this paper, we introduced the \textit{squid} glyph, a 3D vector field uncertainty glyph that accurately represents magnitude and direction uncertainties and is less prone to rotational ambiguity. In addition, we presented a visualization tool that utilizes the \textit{squid} glyph, vector depth, and other techniques for exploration and analysis of vector field uncertainty. 
The \textit{squid} glyph with pointy tip is suboptimal in cases with extensive angular variation. In the future, we plan to investigate the use of tensor glyphs and alternative design strategies, such as arrowheads with less accentuated shapes like half-ellipses to represent vector ensembles with multiple modes and wide variations to further enhance glyph design. 

\acknowledgments{
This work was partially supported by the Intel OneAPI CoE, the Intel Graphics and Visualization Institutes of XeLLENCE, and the DOE Ab-initio Visualization for Innovative Science (AIVIS) grant 2428225. Sandia National Laboratories is a multimission laboratory managed and operated by National Technology and Engineering Solutions of Sandia, LLC, a wholly owned subsidiary of Honeywell International Inc., for the U.S. Department of Energy’s National Nuclear Security Administration under contract DE-NA0003525.
}
\newpage
\bibliographystyle{abbrv-doi}

\bibliography{references}
\end{document}